\documentclass[twocolumn,english,aps,prb,showpacs,superscriptaddress,floats,amsmath,amssymb,floatfix]{revtex4}

\usepackage{color}
\usepackage{amsmath}
\usepackage{graphicx}
\usepackage{graphics}
\usepackage[pdf]{pstricks}
\usepackage{pdftexcmds}
\usepackage{ifpdf}
\usepackage{amssymb}
\usepackage{esint}
\usepackage{comment}

\makeatletter
\@ifundefined{textcolor}{}
{%
 \definecolor{BLACK}{gray}{0}
 \definecolor{WHITE}{gray}{1}
 \definecolor{RED}{rgb}{1,0,0}
 \definecolor{GREEN}{rgb}{0,1,0}
 \definecolor{BLUE}{rgb}{0,0,1}
 \definecolor{CYAN}{cmyk}{1,0,0,0}
 \definecolor{MAGENTA}{cmyk}{0,1,0,0}
 \definecolor{YELLOW}{cmyk}{0,0,1,0}
 }

\allowdisplaybreaks
\makeatother

\begin{document}

\title{Long range interactions in antiferromagnetic quantum spin chains}

\author{B.\ Bravo}
\affiliation{IFLP-CONICET and Departamento de F\'{\i}sica, Universidad Nacional de La Plata, C.C.\ 67, 1900 La Plata, Argentina}

\author{D.C.\ Cabra}
\affiliation{IFLP-CONICET and Departamento de F\'{\i}sica, Universidad Nacional de La Plata, C.C.\ 67, 1900 La Plata, Argentina}

\author{F.A.\ G\'omez Albarrac\'{\i}n}
\affiliation{IFLP-CONICET and Departamento de F\'{\i}sica, Universidad Nacional de La Plata, C.C.\ 67, 1900 La Plata, Argentina}

\author{G.L.\ Rossini}
\affiliation{IFLP-CONICET and Departamento de F\'{\i}sica, Universidad Nacional de La Plata, C.C.\ 67, 1900 La Plata, Argentina}

\date{\today}
\begin{abstract}
We study the role of long range dipolar interactions on antiferromagnetic spin chains, 
from the classical $S\to \infty$ limit to the deep quantum case $S=1/2$, including a transverse magnetic field. 
To this end, we combine different techniques such as classical energy minima, classical Monte Carlo, 
linear spin waves, bosonization and DMRG. 
We find a phase transition from the already reported  dipolar ferromagnetic region to an antiferromagnetic region 
for high enough antiferromagnetic exchange.
Thermal and quantum fluctuations destabilize the classical order before reaching magnetic saturation in both phases, and also close to zero field in the antiferromagnetic phase. 
In the extreme quantum limit $S=1/2$, extensive DMRG computations show that the main phases remain present with transition lines to saturation significatively shifted to lower fields, in agreement with the bosonization analysis. The overall picture keeps close analogy with the phase diagram of the anisotropic XXZ spin chain in a transverse field.
\end{abstract}


% 75.10.Pq Spin chain models
% 75.10.Jm Quantized spin models, including quantum spin frustration
% 75.30.Kz Magnetic phase boundaries (including classical and quantum magnetic transitions, 
% 75.30.Gw Magnetic anisotropy
% 05.30.Rt Quantum phase transitions

\maketitle

\section{Introduction}

Long range interactions in quantum systems have recently attracted much attention.
% in the context of Rydberg atoms,  optical lattices, etc.
While short-range interactions are naturally present in quantum gases, longer-range interactions are much harder to control. 
To investigate them, ultracold gases of particles with large magnetic or electric dipole moments, atoms in Rydberg states, 
or cavity-mediated interactions have been studied.\cite{Landig2016}
These experiments also open the possibility of simulating dipole-dipole interactions in one dimensional spin chains.\cite{Kestner2011}. Since then, theoretical and numerical investigation of dipolar spin chains has been revitalized.\cite{Isidori2011}  
On the other hand the inclusion of dipolar, and more generally long-ranged interactions, in classical and quantum models has proven to modify in different degrees the outcoming physics \cite{Pich1997,Zhang2011,Gessner2016}.  

Motivated by these studies, 
we consider the competition between short range antiferromagnetic exchange and  
long range dipolar interactions in spin chains in order to explore the presence of novel phases, either ordered or disordered.
We also include an external magnetic field, transverse to the dipole-dipole induced anisotropy, 
which competes with both antiferromagnetic and dipolar classical ordering.

The present analysis follows different approaches that allow to explore from  
the classical $S\to \infty$ limit to the deep quantum case $S=1/2$, 
as well as parameter regions where antiferromagnetic exchange dominates over dipolar interactions 
and the other way around. 
For  large $S$ we perform classical energy analysis, classical Monte Carlo simulations 
and linear spin wave fluctuations.
On the other extreme, for spin $S=1/2$ we employ bosonization techniques and DMRG computations. 

With these techniques, we first draw a semiclassical phase diagram showing three distinctive phases: 
a dipole-induced ferromagnetic phase, 
a disordered Luttinger liquid phase and 
an antiferromagnetic Ising N\'{e}el phase. 
The stability of the boundaries between those phases and their transition to the magnetically saturated phase 
is discussed in terms of thermal and quantum spin wave fluctuations. 
Beyond fluctuations, for the $S=1/2$ deep quantum case 
our main results show  a substantial reduction of the magnetic field 
rendering both the dipolar and antiferromagnetic ordered phases into the magnetically saturated phase.
The quantum phases found are in correspondence with those present in an anisotropic  $S=1/2$ spin chain; 
in this sense the main effect of dipolar interactions seems to be encoded in their nearest neighbors contribution.  

The manuscript is organized as follows. In Section II  we define the model and review the classical magnetic phases.  
In Section III we discuss the effect of thermal and quantum  fluctuations by means of Monte Carlo simulations and spin wave calculations.
Our main results for spin $S=1/2$ are presented in Section IV through extensive DMRG data, preceded by a bosonization analysis.  
We close this work with concluding remarks in Section V.

%%%%%%%%%%%%%%%%%%%%%%%%%%%%%%%%%%%%%%%%%%%%%%%%%%%%%%

\section{Model and classical description} \label{sec:class}

We consider a spin chain in the $x$ direction with long range dipolar interactions and nearest neighbours antiferromagnetic  exchange coupling $J>0$. The
%corresponding  
Hamiltonian reads

\begin{eqnarray} \label{eq:H}
 \mathcal{H} &=& J\sum_{i} \vec{S}_i\cdot\vec{S}_{i+1} 
+ \mu^2\sum_{i<j}\left(\frac{\vec{S}_i\cdot\vec{S}_j}{|\vec{r}_i-\vec{r}_j|^3} - 3\frac{S_i^xS_j^x}{|\vec{r}_i-\vec{r}_j|^3}\right) 
\nonumber \\
 & - & h\sum_iS_i^z 
\end{eqnarray}

\noindent where $\vec{S}_i$ is a dimensionless spin $S$ operator at site $\vec{r}_i=i\, a\breve{x}$ with $a$ is lattice spacing between consecutive spins and $\mu$ is the effective gyromagnetic moment of the spins. Since $|\vec{r}_i-\vec{r}_j|=a|i-j|$ we will use in the following a dipolar coupling $D=\frac{\mu^2}{a^3}$ with units of energy. An external magnetic field $\vec{h}=h\breve{z}$  is chosen to be perpendicular to the anisotropy introduced by dipolar interactions along the spin chain direction.

\begin{figure}[ht] 
 \includegraphics[width=0.7 \columnwidth]{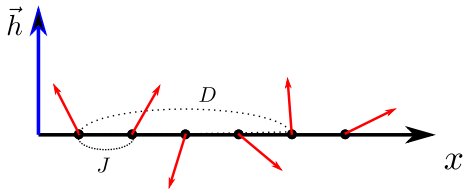}
 \caption{\label{fig:cartoon} 
 (color online) Schematic representation of nearest neighbours exchange interactions and long range dipolar interactions. Transverse magnetic field is also depicted.}
\end{figure}

In this section we review the classical ground configurations of the Hamiltonian in Eq.~(\ref{eq:H}). Dipolar interactions tend to align classical spins along the $x$ axis while the transverse magnetic field induces a tilting towards the $z$ axis. Ferromagnetic exchange couplings $J<0$ are fully satisfied in such configurations \cite{Isidori2011}, but the present antiferromagnetic couplings $J>0$ would prefer to set a N\'{e}el order in the $xy$ plane. One can then search for the classical ground state configurations within the manifold parameterized by
\begin{equation} \label{eq:Ansatz}
\vec{S}_i=S(\sin\theta \cos\phi,(-1)^i \sin\theta \sin\phi,\cos\theta)
\end{equation}
where $\theta$ is the tilting angle w.r.t.\ the magnetic field and $\phi$ describes a staggered deviation from the chain direction into the $xy$ plane. The energy per site of such configurations reads
\begin{eqnarray} 
\label{eq:classical energy} 
\frac{E_0}{S^2} & = &J \left(\cos^2\theta +\sin^2\theta \cos(2\phi)\right) \nonumber \\
& & + \zeta(3)D\left(\cos^2\theta - 2 \sin^2\theta \cos^2\phi -\frac{3}{4}\sin^2\theta \sin^2\phi\right) \nonumber \\
& & -\frac{h}{S} \cos\theta
\end{eqnarray}
where $\zeta(3)=\sum_{n=1}^{\infty} \frac{1}{n^3} \approx 1.20206$ stems from the numerical evaluation of the Riemann zeta function. 

\begin{figure}[ht] 
 \includegraphics[width=\columnwidth]{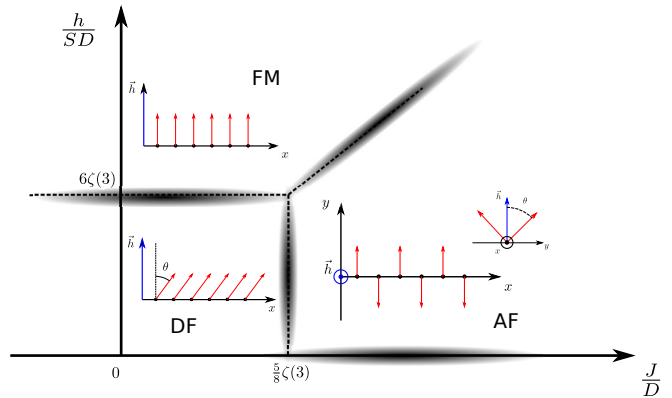}
 \caption{\label{fig:classical-phase-diagram} 
 (color online) 
 Classical phases are described in the $h$ vs.\ $J$ plane: tilted dipolar-ferromagnetic (DF), 
 tilted antiferromagnetic (AF) and magnetic saturation (FM).  
 Shaded regions indicate predominance of both thermal and spin wave quantum fluctuations. 
We show below that the boundaries to saturation are significantly shifted down in the quantum $S=1/2$ limit. 
}
\end{figure}

It is readily found that this classical energy, for $J\leq J_c \equiv \frac{5}{8} \zeta(3) D$ 
and below a saturation field $h_{sat}^\text{DF}(D)=6 \zeta(3) S D $,
exhibits a minimum at $\phi=0$ (parallel spins in the $xz$ plane) and tilting angle
\begin{equation}
 \theta^\text{DF}(h,D)=\arccos\left( \frac{h}{6 \zeta(3) S D} \right) 
\end{equation}
defining a "dipolar-ferromagnetic" (DF) phase. This phase has a ${\mathbb Z}_2$ mirror degeneracy under exchange $\theta \to -\theta$.
Instead, for $J\geq J_c$  and below a saturation field $h_{sat}^\text{AF}(J,D)=4S(J+\frac{7}{8}\zeta(3) D)$,  
the minimum appears at $\phi=\pm \pi/2$ and
\begin{equation}
 \theta^\text{AF}(h,J,D)=\arccos\left(\frac{h}{4S(J+\frac{7}{8}\zeta(3) D)}\right)
\end{equation}
defining an antiferromagnetic (AF) phase where 
spins lie in staggered tilted directions in the $yz$ plane. 
This phase has a ${\mathbb Z}_2$ discrete translation degeneracy along the chain direction.
The classical phase diagram in the $h$ vs $J$ plane is shown in Fig.~\ref{fig:classical-phase-diagram}.

At the specific value $J=J_c>0$  the ground state manifold is infinitely 
degenerate for any value of the external magnetic field below saturation, 
with a free value for the angle $\phi$  that can continuously 
interpolate between the dipolar-ferromagnetic and the antiferromagnetic phases. 
Also notice that at zero field the classical AF phase possesses
an extra $U(1)$ degeneracy associated to rotations around the chain direction. 
Such degeneracies enhance the role of classical and quantum fluctuations, as we discuss in the next Section.

%%%%%%%%%%%%%%%%%%%%%%%%%%%%%%%%%%%%%%%%%%%%%%%%%%%%%%

\section{Classical and Quantum Fluctuations} \label{sec:fluctuations}

%%%%%%%%%%%%%%%%%%%%%%%%%%%%%%%%%%%%%%%%%%%%%%%%%%%%%%

% algo? 

\subsection{Monte Carlo simulations}

We study here  the effects of classical thermal fluctuations on the model presented in Section \ref{sec:class}, 
in order to check the stability of the phase boundaries of the zero temperature classical phase diagram shown 
in Fig.~\ref{fig:classical-phase-diagram}.  
To this end we have run Monte Carlo simulations using the standard Metropolis algorithm combined with overrelaxation (microcanonical) updates 
for chains of $L=300$ sites. 
Finite size systems with periodic boundary conditions are considered, so that long range interactions are taken up to $L/2$ neighbors. 
For each simulation $10^4$ Monte Carlo steps (mcs) were dedicated to thermalization, 
lowering the temperature with the annealing technique at a rate $T_{n+1}=0.9\times T_n$.
Measurements are then taken during the $2\times10^4$ subsequent mcs.  
The results presented for each data point describe the average over 100 independent simulations.

In order to describe the effects of thermal fluctuations in both DF phase, $J<J_c$  and AF phase, $J>J_c$, 
we define two different order parameters: a mean magnetization $m$ defined as
\begin{equation}
m = \sqrt{m_x^2+m_y^2+m_z^2}  \label{eq:MF}
\end{equation}
where $m_\alpha=\frac{1}{L}\sum_i S^\alpha_i$ $(\alpha=x,y,z)$ and an antiferromgnetic magnetization $m^\text{AF}$ defined as
\begin{equation}
m^\text{AF} = \sqrt{m_x^2+(m_y^{stag})^2+m_z^2} \label{eq:MAF}
\end{equation}
where $m_y^{stag}=\frac{1}{L}\sum_i (-1)^iS^y_i$ picks the staggered contribution of the $y$ components.

\begin{figure}[ht] 
\includegraphics[width=0.95\columnwidth]{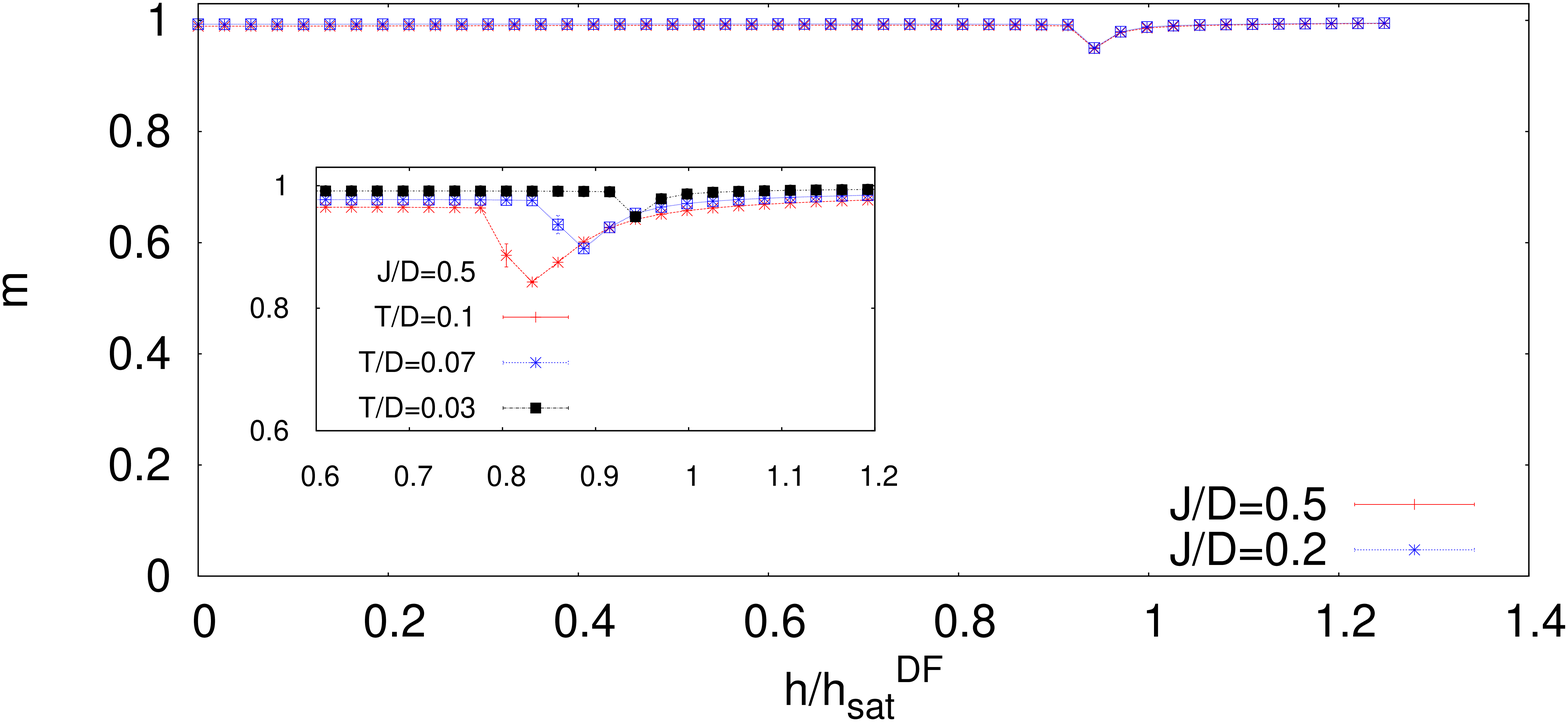}
\includegraphics[width=0.95\columnwidth]{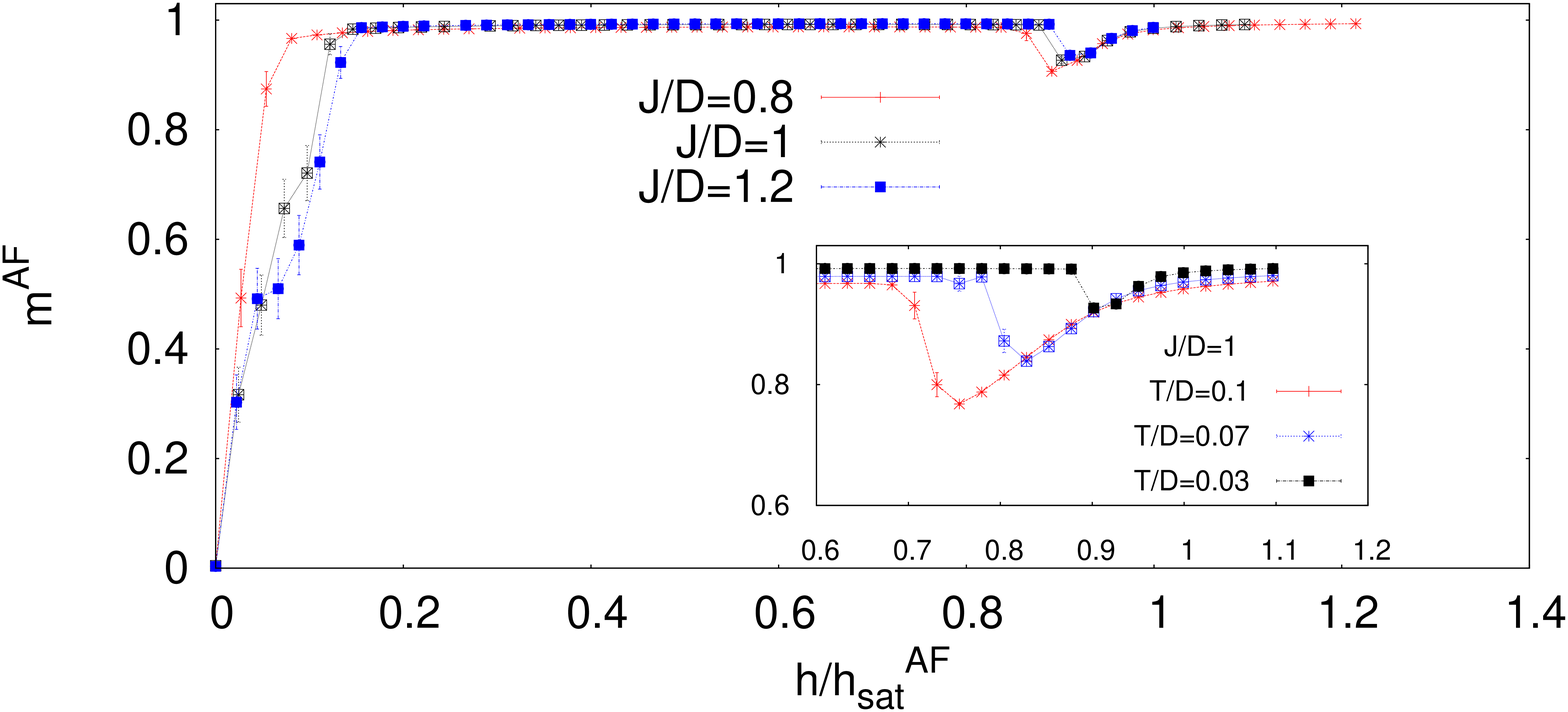}
\caption{\label{fig:classicMvsh} (color online) Order parameters $m$ (top) and $m^\text{AF}$ (bottom) 
as functions of the external magnetic field scaled with the corresponding $T=0$ saturation values. 
Temperature is set at $T/D=0.02$ for different values of $J/D$ for $J<J_c$ (top) and  $J>J_c$ (bottom). }
\end{figure}

We show $m$ and $m^\text{AF}$ as functions of 
the external magnetic field
%$h/h_{sat}^\text{DF,AF}$ 
for different values of $J$ and $T/D$ in Fig.~\ref{fig:classicMvsh}.
The top panel shows $m$ for two values of $J<J_c$ at $T/D=0.02$. The external magnetic field is normalized by the corresponding $T=0$ saturation value $h_{sat}^\text{DF}$ 
(see section \ref{sec:class}). There is a clear dip at $h/h_{sat}^\text{DF} < 1$ for all values of $J$, due to temperature effects.
This is illustrated in the inset of the top panel, which shows $m$ vs $h/h_{sat}^\text{DF}$ for $J/D=0.5$ at different temperatures. 
Analogously, the bottom panel shows $m^\text{AF}$ versus $h/h_{sat}^\text{AF}$ for three values of $J>J_c$ and $T/D=0.02$, while the inset shows the effects of temperature for $J/D=1$.

A dip close to saturation is observed, as expected, in both DF and AF phases which increases with temperature.
In the AF phase, on the other hand, one observes a much pronounced dip in the staggered magnetization close to zero field due to the $U(1)$
symmetry of the AF classical solution in absence of magnetic field. 

\subsection{Spin Waves spectrum}

Linear spin waves (LSW) fluctuations around the classical $S\to \infty$ solutions can be analyzed in a standard way. 
One first introduces local axes such that at each site $i$ a new $z'$ axis coincides with the classical solution spin orientation.
Within the LSW approximation, the spin operator components in the local axes can be represented by bosonic 
Holstein-Primakoff local operators $a_i$, $a^\dagger_i$ as
\begin{eqnarray}
 S^{z'}_i&=&S-a^\dagger_i a_i\\
 S^{x'}_i&=&\sqrt{\frac{S}{2}}(a^\dagger_i + a_i) \nonumber \\
 S^{y'}_i&=&i\sqrt{\frac{S}{2}}(a^\dagger_i - a_i) \nonumber
\end{eqnarray}
After a Fourier transformation in a periodic chain with $L$ sites,
\begin{equation}
 a_i=\frac{1}{\sqrt{L}}\sum_k e^{ikx_i} a_k  
\end{equation}
where $k=p \frac{2\pi}{La}, \,p=-L/2,\cdots,L/2$ and $x_i=i\,a$, and ignoring cubic and higher order terms in $a_k$, $a^\dagger_k$, one can get the form
\begin{equation}
 {\mathcal H}=S\sum_k \left\lbrace A_k(a^\dagger_{k} a_{k} + a^\dagger_{-k} a_{-k})+B_k(a^\dagger_{k} a^\dagger_{-k} + a_{k} a_{-k})\right\rbrace
 \label{eq:generic_anomalous_H}
\end{equation}
with real $A_k \geq |B_k|$. Then the Hamiltonian can be diagonalized by a Bogoliubov transformation
\begin{eqnarray}
 a_{k} &=& u_k \alpha_k +v_k \alpha^\dagger_{-k} \nonumber \\
 a^\dagger_{-k} &=& v^*_k \alpha_k +u^*_k \alpha^\dagger_{-k} \label{eq:Bogo}
\end{eqnarray}
with $|u_k|^2-|v_k|^2=1$ ensuring that $\alpha_k$, $\alpha^\dagger_k$ are bosonic modes. The Hamiltonian finally reads 
\begin{equation}
 {\mathcal H} = N E_0 + \frac{S}{2}\sum_k \left(\varepsilon_k - A_k\right) + S \sum_k \varepsilon_k \alpha^\dagger_k \alpha_k
 \label{eq:H_Bogo}
\end{equation}
where $E_0$ is the classical energy per site given in Eq.~(\ref{eq:classical energy}) and 
$\varepsilon_k=\sqrt{A_k^2-|B_k|^2}$ are the Bogoliubov mode energies.

At the dipolar-ferromagnetic phase an appropriate global orthogonal coordinate system is set by rotating the original axes 
an angle $\theta_\text{DF}(h,D)$ around the $y$ direction. 
The Hamiltonian coefficients $A^\text{DF}_k$, $B^\text{DF}_k$ in Eq.~(\ref{eq:generic_anomalous_H}) are given by
\begin{eqnarray}
 A^\text{DF}_k &=& J \left(\cos(ka)-1\right) + \nonumber \\
 & & +D\left( 3\sin^2\theta^\text{DF}-1\right)\left(\zeta(3)+\frac{1}{2}F(ka) \right) + \frac{h}{2S} \cos\theta^\text{DF} \nonumber \\
 B^\text{DF}_k &=& -\frac{3}{2}D \cos^2 \theta^\text{DF}F(ka)
 \label{eq:coeff_F}
\end{eqnarray}
where $F(ka)\equiv Re[Li_3(e^{ika})]$ with $Li_3(z)$ the polylogarithm\footnote{$F(ka)$ 
is also known as the Clausen function $Cl_3(ka)$.} series $Li_3(z)\equiv \sum_{n=1}^\infty \frac{1}{n^3}z^n$.

At the antiferromagnetic phase the appropriate axes are local ones, obtained by rotating the original axes at each site $i$ with 
an angle $(-1)^i \theta^\text{AF}(h,J,D)$ about the $x$ direction.
The Hamiltonian coefficients $A^\text{AF}_k$, $B^\text{AF}_k$ in Eq.~(\ref{eq:generic_anomalous_H}) are then given by
\begin{eqnarray}
 A^\text{AF}_k &=& J \cos^2 \theta^\text{AF} \cos(ka) - J\cos(2 \theta^\text{AF}) +\nonumber \\ 
 & &  +D(\frac{1}{2} \cos^2 \theta^\text{AF} - 1) F(ka) + \frac{1}{2} D \sin^2 \theta^\text{AF} G(ka) + \nonumber \\
 & &  -D\zeta(3) (\cos^2 \theta^\text{AF}-\frac{3}{4}\sin^2 \theta^\text{AF})+\frac{h}{2S} \cos \theta^\text{AF}\nonumber \\
 B^\text{AF}_k &=& J \sin^2 \theta^\text{AF} \cos(ka) -D\left(\frac{1}{2} \cos^2 \theta^\text{AF} + 1\right) F(ka)+\nonumber \\
 & & -\frac{1}{2} D \sin^2 \theta^\text{AF} \,G(ka)
 \label{eq:coeff_AF}
\end{eqnarray}
where $G(ka)\equiv\frac{1}{4}F(2ka)-F(ka)$.

In either case, the semiclassical ground state is the Bogoliubov vacuum $|0\rangle$ annihilated by the operators $\alpha_k$. The ground state energy is then given by 
\begin{equation}
 {\mathcal H} = N E_0 + \frac{S}{2}\sum_k \left(\varepsilon_k - A_k\right) .
 \label{eq:E_semiclassical}
\end{equation}

Within the LSW framework the sensible order parameter to compute is the average of the local magnetizations along the classical directions $z'$, defined as 
\begin{equation}
 m^{z'}=\frac{1}{N}\sum_i \langle 0|S_i^{z'}|0 \rangle = S+\frac{1}{2}-\sum_k \frac{A_k}{2\sqrt{A_k^2-|B_k|^2}}.
\end{equation}
A large value of the summation in the last term signals the breakdown of the LSW approximation, meaning that quantum 
fluctuations destroy the classical order. 
This occurs in the shaded regions of the classical phase diagram in Fig.~\ref{fig:classical-phase-diagram}. 

\begin{figure}[ht] 
 \includegraphics[width=\columnwidth]{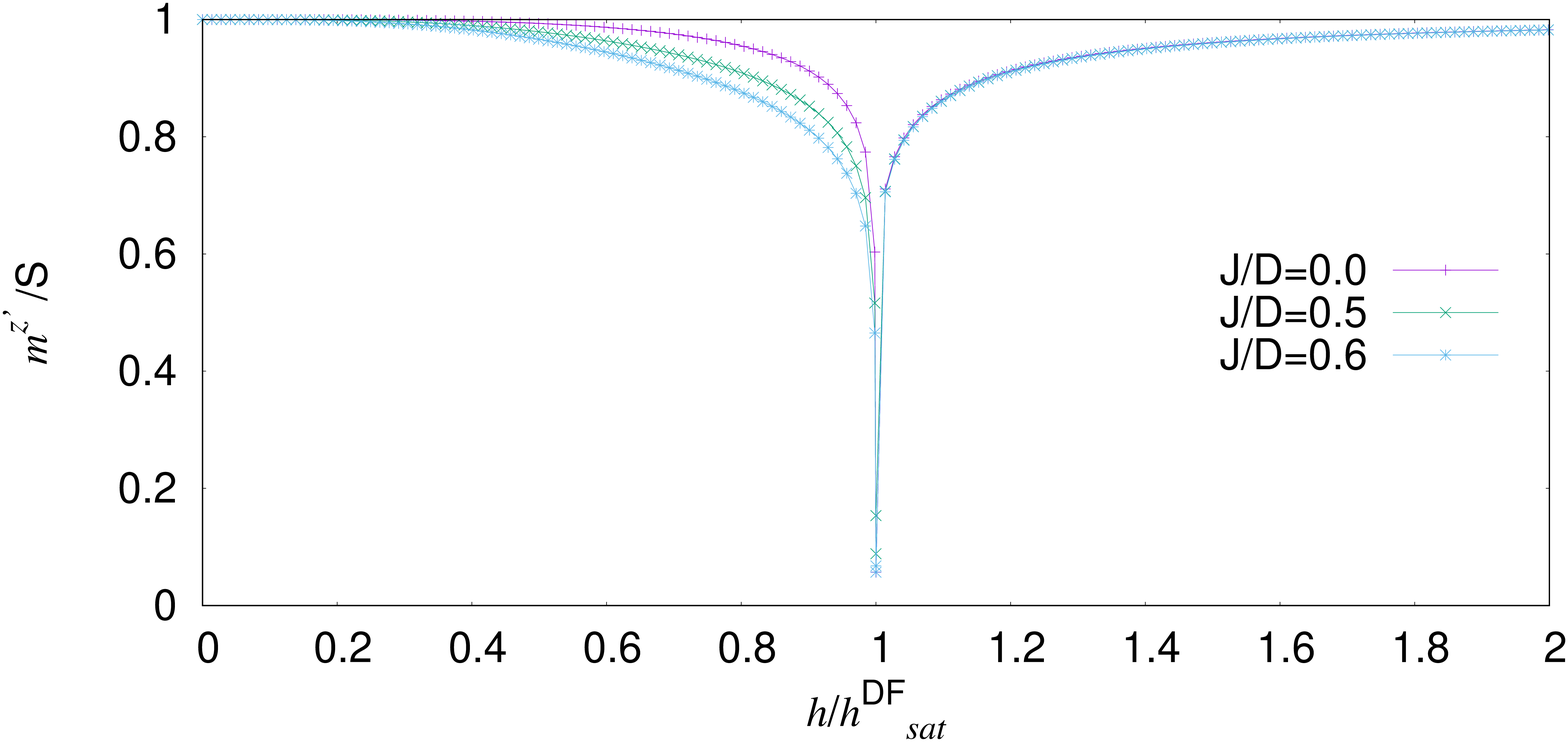}
 \caption{\label{fig:mlsw-ferro} 
 (color online) Mean local magnetization vs. $h$ for different values of antiferromagnetic  $J<J_c$, in the LSW approximation.}
\end{figure}

\begin{figure}[ht] 
 \includegraphics[width=\columnwidth]{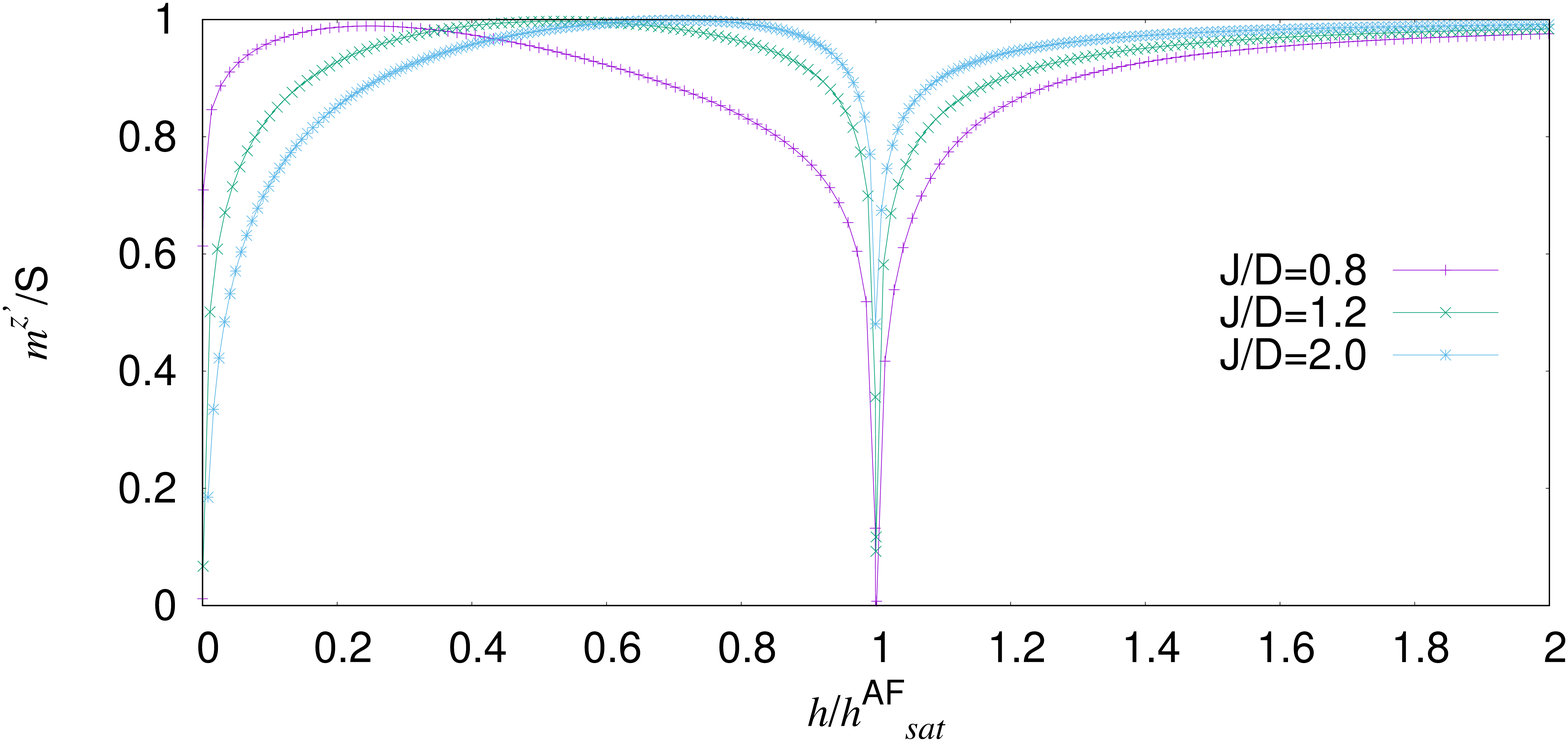}
 \caption{\label{fig:mlsw-antiferro} 
 (color online) Mean local magnetization vs. $h$ for different values of antiferromagnetic  $J>J_c$, in the LSW approximation. }
\end{figure}

We show in Fig.~\ref{fig:mlsw-ferro}  $m^{z'}$ vs.\ $h$ for coupling ratios within the DF phase, $J/D < \frac{5}{8} \zeta(3)$, 
while in Fig.~\ref{fig:mlsw-antiferro} we show the corresponding results for ratios within the AF phase, $J/D > \frac{5}{8} \zeta(3)$. 
These results extend those obtained in Ref.~\onlinecite{Isidori2011}, at $J=0$, 
to the whole DF region and show up a new phase with antiferromagnetic characteristics. 
One can observe that the LSW corrections to the order parameter diverge as $h$ approaches $h_\text{sat}$, 
both in the DF and the AF phases. 
From the slopes in these figures the sensitivity to quantum fluctuations shows to be higher when $J$ approaches $J_c$,
{\em i.e.} when the competition between dipolar and exchange interactions is stronger. 
In the AF phase quantum fluctuations are crucial not only close to saturation but also close to zero 
field, as expected for antiferromagnetic systems. 
Here the LSW corrections are more important for larger $J$. Indeed, the crossings in the curves in Fig.~5 signal a crossover from a dipolar reminiscent behavior, close to $J_C$, towards an exchange dominated behavior for $J \gg D$.

These features agree with the classical thermal fluctuations picture and are confirmed by extensive DMRG computations and a bosonization analysis in the next Section.

%%%%%%%%%%%%%%%%%%%%%%%%%%%%%%%%%%%%%%%%%%%%%%%%%%%%%%

%%%%%%%%%%%%%%%%%%%%%%%%%%%%%%%%%%%%%%%%%%%%%%%%%%%%%%%%%%%%%%%%%%%%%%%%%%%%%%%%%%%%%%%%%%%%%%%%%%
\section{Extreme quantum limit: spin $S=1/2$ case}

In this Section we discuss specific features of the spin $S=1/2$ case, 
using bosonization techniques and extensive state of the art  DMRG numerical computations for long range coupled systems.

\subsection{Bosonization approach}

For the present discussion we find it convenient to rewrite the Hamiltonian in Eq.~(\ref{eq:H}) by 
separating the nearest neighbors interactions ($\mathcal{H}_\text{NN}$)
from the longer distance dipolar terms ($\mathcal{H}_\text{int}$), which we then treat perturbatively. For the sake of clarity, 
we consider first the zero field case. The Hamiltonian reads
\begin{equation} \label{eq:H-separate}
  \mathcal{H} = \mathcal{H}_\text{NN} + \mathcal{H}_\text{int}
\end{equation}
where
\begin{equation} \label{eq:H-NN}
 \mathcal{H} = (J+D)\sum_i \left[S_i^y S_{i+1}^y + S_i^z S_{i+1}^z+ \frac{J-2D}{J+D} S_i^x S_{i+1}^x \right] 
\end{equation}
and
\begin{equation} \label{eq:H-int}
\mathcal{H}_\text{int}= D\sum_{j>i+1}\left(\frac{-2 S_i^x S_j^x + S_i^y S_j^y + S_i^z S_j^z}{(j-i)^3} \right) .
\end{equation}
Notice that $\mathcal{H}_\text{NN}$ corresponds to the well known anisotropic XXZ Heisenberg chain 
with $\Delta=\frac{J-2D}{J+D}$ (see for instance Ref.~\onlinecite{Mikeska2004}). 
In this sense, the short range effect of dipolar interactions is the onset of an exchange-like anisotropy along the chain direction.
Without further interactions, a Luttinger liquid phase 
%(also called XY phase) 
is present, extending from a  ferromagnetic transition point at $J= 0.5 D$ ($\Delta=-1$) 
up to an isotropic Heisenberg point ($\Delta=1$) to be reached at $J/D \to  \infty$, 
with a free fermion point ($\Delta=0$) located at $J=2 D$.
Bosonization of the spin $S=1/2$ chain in this regime has succeeded to describe the ground state 
and correlation functions, 
as well as allowing for a conformal perturbative scheme
(see for instance Ref.~\onlinecite{Giamarchi2004}). 
Following this scheme one can show that the $1/r^3$  decaying long range dipolar interactions 
in $\mathcal{H}_\text{int}$ do not alter the Luttinger liquid behavior, 
but only renormalize the Luttinger parameters. 
In accordance with the spin wave indications, 
the system enters a disordered phase for $J>J^q_c$, 
a quantum critical point which is eventually shifted from $0.5 D$ by dipolar interactions. 

In the region  $J< 0.5 D$ the truncated Hamiltonian $\mathcal{H}_\text{NN}$ enters 
a gapped ferromagnetic phase \cite{Mikeska2004}, 
with a two-fold degenerate ground state 
characterized by the order parameter $\langle S^x_i\rangle \neq 0$. 
Though conformal perturbations cannot be applied in this region, 
notice that the full Hamiltonian $\mathcal{H}_\text{NN}+\mathcal{H}_\text{int}$ 
classically exhibits the same ordering, which is robust against thermal and quantum fluctuations.
For the quantum case $S=1/2$ we explore numerically this order parameter in the next section.

The behavior of the present system under a transverse magnetic field, 
both in the Luttinger and the ferromagnetic region,
can be naturally related to the anisotropic XXZ Heisenberg chain in a transverse field. 
As discussed in Refs.~\onlinecite{Dmitriev2002} and \onlinecite{Caux2003},
a small transverse field induces a gap on the Luttinger phase, 
driving the spin system into an Ising N\'eel phase with staggered 
non vanishing expectation value of spin components along the $y$ direction ({\em i.e.}\ transverse to both 
the anisotropy and the magnetic field). 
On the gapped ferromagnetic phase, 
the transverse field diminishes the order parameter, until a quantum phase transition into a paramagnetic phase is reached.
Discussing how the longer range interactions in $\mathcal{H}_\text{int}$ may modify this picture is beyond
the scope of the present paper. 

These predictions, namely the renormalization of the quantum critical point and 
the behavior in a transverse magnetic
field,  are confirmed below by a DMRG analysis.

\subsection{DMRG calculations}

With the aim of characterizing the spin $S=1/2$ quantum phases of the antiferromagnetic chain with dipolar interactions, we have analyzed the ground state of the Hamiltonian in Eq.~(\ref{eq:H})  with the DMRG technique \cite{White}. Including long range interactions is a non-trivial task, comparable with current studies of two-dimensional spin systems \cite{DMRG2D,Bravo-Gazza}. 
Based on our experience in such investigations,  we use here open boundary conditions and long-range interactions 
involving all available neighbors in chains of finite size. 
We have considered chains of length $L$ up to $64$ sites, keeping $m=350$ states and achieving truncation errors in the density matrix of the order of 
$10^{-11}$.

\begin{figure} 
 \includegraphics[width=\columnwidth]{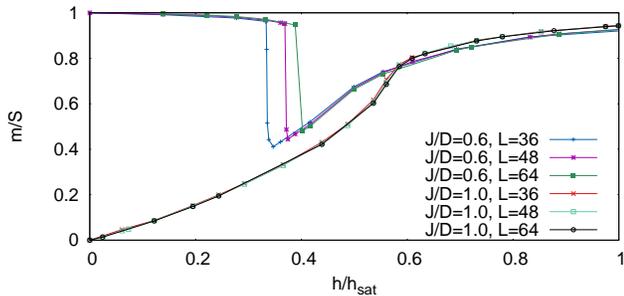}
 \caption{\label{fig:total_magnetization} (color online) Total magnetic moment $m/S$ as a function of the applied magnetic field $h$ 
 for two representative values of the antiferromagnetic exchange $J/D = 0.6 <J_c$ and $J/D = 1.0 >J_c$ and lattice sizes $L$=36,48,64
 (each curve is normalized by its corresponding saturation magnetic field $h_{sat}^{DF}$ and $h_{sat}^{AF}$).
 } 
\end{figure}

As a first step to identify the configurations of the system we compute the total magnetic moment $m$ defined in Eq.~(\ref{eq:MF}),
% \begin{equation}
% m = \sqrt{m_x^2+m_y^2+m_z^2} 
% \end{equation}
where now $m_\alpha = \frac{1}{L}\sum_{i} \langle S^\alpha_i \rangle$ stands for the average of quantum expectation values.
In Fig.~\ref{fig:total_magnetization} we show DMRG results for the total magnetic moment $m$ as a function of the magnetic field
for two representative values $J=0.6\, D$ and $J=1.0\, D$ of the antiferromagnetic exchange interaction, chosen 
to be compared  with the figures shown in Section III.  
As in the classical chain, we observe two different configurations depending on the value of the exchange interaction
$J$ being below or above a critical value. In more detail (not shown in the figure) we have been able to estimate
a quantum critical coupling $J_c^q=0.7275(25)$  slightly lower than the classical value obtained in Section \ref{sec:class}, $J_c=\frac{5}{8} \zeta(3) D\approx 0.7512$.
In the DF phase, $J<J_c^q$, the magnetization exhibits a minimum,  
a similar behavior to that of the pure dipolar case in Ref.~\onlinecite{Isidori2011}.
Instead, in the AF phase, $J>J_c^q$, we find that the total magnetization smoothly increases with the 
applied magnetic field. 

As a further step we analyze the components of the magnetic moments 
along the three directions $x$, $y$ and $z$ separately.
\begin{figure}[htp!] 
 \includegraphics[width=\columnwidth]{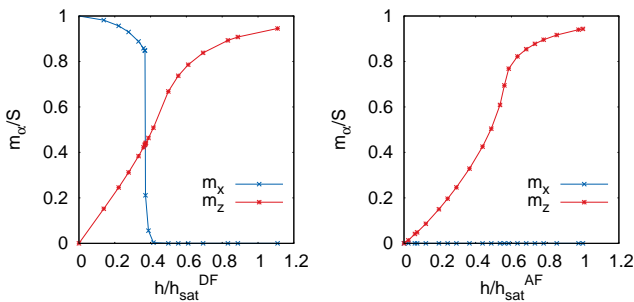}
 \caption{\label{fig:mx_mz} 
 (color online) Magnetic moments $m_x$ and $m_z$ as a function of the applied magnetic field for a chain of $L=48$ sites. 
 Left panel: Results for $J/D=0.6$ ($J<J_c$). Right panel: Results for $J/D=1.0$ ($J>J_c$).
%  Left panel: Magnetic moments $m_x$ and $m_z$ for $J/D=0.6$ ($J<J_c$) as a function of the magnetic field. 
%  Right panel: Magnetic moments $m_x$ and $m_z$ for $J/D=1.0$ ($J>J_c$). 
%   (L=48) 
  } 
\end{figure}
In Fig.~\ref{fig:mx_mz} we show the values of $m_x$ and $m_z$ as functions of the applied magnetic field, 
for exchange $J=0.6\,D<J_c^q$ and $J=1.0 \,D>J_c^q$. 
In both phases $m_z$ increases smoothly with the applied field, approaching to saturation at about the classical saturation field. Instead, the $m_x$ component makes apparent the difference between the DF and AF regimes. 
The component $m_y$ vanishes in both cases, and for the sake of clarity it is not shown in the figures; 
we argue below that the reason why this happens is very different for each phase.

In the DF phase ($J<J_c^q$) we found a two-fold degenerate ground state, as dictated by parity ${\mathbb Z}_2$ symmetry under $x \to -x$ reflections. 
A parity resolved basis for this ground state subspace is given by states with $m_x<0$ and $m_x>0$. 
A tiny magnetic field  $h_x=10^{-10} D$ 
acting just on the end sites of the chain\cite{Isidori2011}  is enough to explicitly break parity $\mathbb{Z}_2$ symmetry, selecting the state with $m_x>0$. This response supports the interpretation that the finite size ground state is a simple superposition of disentangled ferromagnetic product states.
Following this strategy we produced the states shown in the left panel of Fig.~\ref{fig:mx_mz}:
at low fields the most important contribution to the magnetization is provided by the 
$m_x$ component, followed by a 
sudden drop 
% of $m_x$, 
well before $m_z$ approaches saturation. 
This explains the pit in Fig.~\ref{fig:total_magnetization}. 
As the magnetic field increases, we found a magnetic field $h_c\approx0.47h_{sat}^{DF}$ 
above which all the magnetization weight is already in the $z$-direction. 
The value of $h_c$ is obtained by extrapolation of the $h_c(L)$ for different chain lengths, 
and signals that all spins already align with the magnetic field at less than half the classical saturation field. 
In this phase, as in the pure dipolar case, 
there is no symmetry reason for the system to choose an orientation in the $y$ direction; 
we accordingly found that $\langle S_i^y\rangle = 0$ at each site.

\begin{figure} 
 \includegraphics[width=\columnwidth]{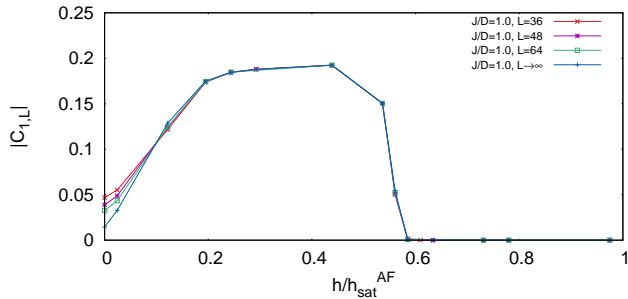}
 \caption{\label{fig:end-to-end_correlations} 
 (color online) End-to-end correlations for lattice lengths $L=36,48,64$ and the $L \to \infty $ extrapolation as a function of
 the applied magnetic field for $J/D = 1.0$.}
% Results of the $L \to \infty $ extrapolation of end-to-end correlations as a function of the applied magnetic field for
% $J/D = 1.0$}
 \end{figure}

In the AF phase ($J>J_c^q$) we observed distinct ground states in the absence or presence of the external magnetic field.
Without external field, in agreement with the Luttinger regime found in the bosonization analysis, the ground state shows no order; 
local expectation values vanish for any spin component as shown in the right panel of Fig.~\ref{fig:mx_mz} at $h=0$. 
Under a magnetic field, also in agreement with the gapped Ising N\'eel phase predicted by bosonization,  
we found a two-fold degenerate ground state, related to one-site ${\mathbb Z}_2$ translation invariance. 
While we expect a spontaneous symmetry breaking in the thermodynamic limit, leading to a
staggered magnetization along the $y$ direction, 
one should notice that translation symmetry is not broken in the available finite size systems\footnote{The argument is valid for periodic boundary conditions. 
Still, one-site translations are an approximate symmetry for the open boundary chains considered in this work}; 
in consequence, one point operators can only show homogeneous expectation values 
and a staggered magnetization is compatible with the observed local result $\langle S_i^y\rangle = 0$.
In contrast to the DF phase, in this case a tiny staggered magnetic field $h_y$ acting on the end sites of the chain is not enough to explicitly break ${\mathbb Z}_2$ translation invariance. % This suggests that the degenerate ground states are highly entangled.
In order to obtain  clear signals of the N\'eel order classically observed  and quantum mechanically predicted in the AF phase, 
we resorted to the computation of $S^y$  two-point spin correlations  $ \langle S_i^y S_j^y \rangle $ which allow for staggered non-vanishing results, invariant under one-site translations.
For $J>J_c^q$ we indeed found staggered correlations, which include an exponentially decaying connected part and a non-vanishing  long range order disconnected part,
% %
\begin{equation}
\langle S_i^y S_j^y \rangle = \langle S_i^y S_j^y \rangle_0 + \langle S_i^y\rangle \langle S_j^y \rangle 
\end{equation}
% %
with $\langle S_i^y\rangle \langle S_j^y \rangle \propto (-1)^{j-i} (m_y^{stag})^2$.
% The correlation decay rate could not be confidently estimated from the available finite size data. 
%
We thus can extract the antiferromagnetic order parameter from the two-point spin correlations.
The long distance behavior can be better analyzed by considering the end-to-end correlations 
\footnote{In order to reduce boundary effects caused by OBC we discarded some sites 
at each end of the chain.} 
\begin{equation}
 C_{1,L} = \langle S_1^y S_L^y \rangle .
\end{equation}
As pointed out in  Ref.~\onlinecite{Fisher1998}  
these correlations between the spins at opposite ends of the chain provide a more tractable distance scaling than bulk correlations.
Because of the even-odd sign of correlators, 
we rather plot $|C_{1,L}|$ to identify the response of end-to-end correlations to the applied magnetic field. 
We plot in Fig.~\ref{fig:end-to-end_correlations} the results for the considered finite size chains 
at $J/D = 1.0$ for several values of the field, 
as well as their infinite size extrapolations.
As a first feature it can be observed that, at intermediate values of the field,  
the order parameter is non zero, with a negligible size scaling. 
This confirms the presence of long range AF order for $J>J_c^q$. 
As a second one, the staggered correlations drop down and vanish at about 60\% of the classical saturation field. 
This coincides with the high slope rise of $m_z$ in the right panel of Fig.~\ref{fig:mx_mz}. 
Finally,  end-to-end correlations approach to zero as the magnetic field vanishes 
(this is more clear deep inside the AF phase, for instance at $J/D=1.5$), 
confirming that staggered order is destroyed at zero field as suggested by 
thermal and quantum fluctuations and symmetry arguments.  
Moreover, the noticeable size scaling indicates the typical quasi long range order in the Luttinger phase. 
These results resemble those found for the Luttinger phase of the anisotropic XXZ spin chain 
in the presence of a transverse field 
(see for instance Fig.~5 in Ref.~\onlinecite{Caux2003}).

% \begin{figure} 
%  \includegraphics[width=\columnwidth]{figs/order_parameter_j1_2.eps}
%  \caption{\label{fig:orderparameter} 
%   Order parameter for several lattice sizes when $J>J_c$. FALTAN REDES MAS GRANDES Y MAS CAMPOS. The inset shows spin correlations $\langle S^y_i S^y_j \rangle$
%   as a function of magnetic field $h$ for a lattice size L=...  }
% \end{figure}

\section{Conclusions}

In the present work we have described the phase diagram for an antiferromagnetic nearest neighbors spin chain 
(with exchange strength $J$)  including the effects of long range dipole-dipole interactions (of strength $D$) 
and a transverse magnetic field (of strength $h$). 

On the one hand, we have characterized the presence of ordered phases for classical spins 
and their stability under the influence of classical 
and quantum fluctuations. 
The main feature here is the presence of a phase transition at $J/D=\frac{5}{8} \zeta(3)$ from a 
dipolar dominated phase  to an antiferromagnetic phase. 
In the former one classical spins are aligned in a ferromagnetic pattern forming an angle  
$\theta^\text{DF}(h,D)=\arccos\left( \frac{h}{6 \zeta(3) S D} \right)$ with the external field, 
according to the competition between the dipolar tendency to align them parallel to the chain 
and the Zeeman energy, 
while in the latter phase spins order antiferromagnetically, transverse to the dipolar anisotropy 
and canted towards the external field at an angle 
$\theta^\text{AF}(h,J,D)=\arccos\left(\frac{h}{4S(J+\frac{7}{8}\zeta(3) D)}\right)$. 
Classical and quantum fluctuations destroy both these classical orders before reaching the saturation field, 
and also the antiferromagnetic order close to zero field.

On the other hand, in the extreme quantum case $S=1/2$, 
DMRG computations indicate that classical order disappears 
well before reaching the classical saturation fields. 
This is  in agreement with previous studies on the purely dipolar $S=1/2$ chain \cite{Isidori2011} 
where a quantum critical point belonging to the 2$d$ Ising universality class
was identified and the effect of quantum fluctuations was proved to reduce the value of the critical field to magnetic saturation.
Regarding the competition between dipolar ferromagnetic and antiferromagnetic orders, 
we have found that the quantum critical point separating these phases 
is slightly shifted (about $3\%$) to a lower value. 
This behavior is fully compatible with the bosonization analysis discussed in the text. 
The observed quantum phases are similar to those present in the well known XXZ anisotropic 
spin chain. 
Also the quantum critical transition reported in Ref.~\onlinecite{Isidori2011} for the dipolar $S=1/2$ chain 
and the corresponding transition line for the XXZ chain belong to the same universality class. 
Our results suggest that the qualitative effect of dipolar interactions 
might be encoded in the anisotropy introduced by  the nearest neighbors terms. 
Were this the case, the shaded phase transition lines in Fig.~2 would be critical. 
Further work along this rationale is in progress and will be reported elsewhere.

The current development of quantum simulations with ultracold trapped atoms is promising for 
testing an ever wider spectrum of quantum many-body properties.
Some key achievements paving the way to the present discussion are the techniques described in Ref.~\onlinecite{Duan2003} 
to simulate spin 1/2 exchange interactions and 
the control of magnetic dipole-dipole interactions between chromium atoms in Ref.~\onlinecite{Stuhler2005},
as well as the stronger electric dipole-dipole interactions between polar molecules in Ref.~\onlinecite{Ni2008}.
Most recently, a controlled competition between short range and long range interactions 
in a bosonic optical lattice 
revealed in Ref.~\onlinecite{Landig2016} the possibility of simulating different quantum phases and their transitions. 
We hope that such techniques could test our predictions in a near future.

\begin{acknowledgments}

BB thanks C.J.\ Gazza for useful help in DMRG implementation details.
FGA acknowledges H.D.\ Rosales and  R.\ Borzi for discussions on Monte Carlo implementation.
DCC and GLR acknowledge M.D.\ Grynberg, A.\ Iucci and R.\ Santachiara for useful discussions.
This work is partially supported by PIP 2015-0813 CONICET and PICT 2012-1724 ANPCyT.

\end{acknowledgments}

\end{document}